\documentclass[hyper,11pt,letterpaper]{JHEP3}
\usepackage[dvips]{epsfig}
\usepackage{amsmath,amssymb,epsf,amsfonts}
\usepackage{graphicx}
\usepackage{dcolumn}
\usepackage{bm}
\usepackage{caption}
\usepackage{subcaption}

\usepackage[english]{babel}
\usepackage{amssymb}
\usepackage{amsfonts}
\usepackage{amsmath}
\usepackage{graphicx}
\usepackage{epsfig}
\usepackage{cite}
\newcommand{\be}{\begin{equation}} \newcommand{\ee}{\end{equation}}
\newcommand{\bea}{\begin{eqnarray}} \newcommand{\eea}{\end{eqnarray}}
\newcommand{\beann}{\begin{eqnarray*}}  \newcommand{\eeann}{\end{eqnarray*}}
\newcommand{\bfig}{\begin{figure}} \newcommand{\efig}{\end{figure}}
\newcommand{\ba}{\begin{array}} \newcommand{\ea}{\end{array}}
\newcommand{\bcen}{\begin{center}} \newcommand{\ecen}{\end{center}}
\newcommand{\btab}{\begin{tabular}} \newcommand{\etab}{\end{tabular}}

\newtheorem{Proposition}{Proposition}[section]

\newtheorem{Theorem}{Theorem}[section]
\newtheorem{Lemma}{Lemma}[section]
\newtheorem{Corrolary}{Corrolary}[section]

\newcommand{\bp}{\begin{Proposition}}   \newcommand{\ep}{\end{Proposition}}
\newcommand{\bt}{\begin{Theorem}}   \newcommand{\et}{\end{Theorem}}
\newcommand{\bl}{\begin{Lemma}}     \newcommand{\el}{\end{Lemma}}
\newcommand{\bc}{\begin{Corrolary}} \newcommand{\ec}{\end{Corrolary}}
\title{
Minimal Submanifolds asymptotic to $AdS_4 \times S^2$ in $AdS_5 \times S^5$
}
\author{
Han-Chih Chang 
\\
Department of Physics, University of Washington, Seattle, WA 98195-1560\\
E-mail: \email{hanchih@uw.edu}
}

\abstract{

In this work, as an attempt in understanding the interplay between
deformations of the external and internal spaces associated with the probe-brane 
embedding submanifold, we construct
the zero-temperature phase diagram for the
coupled phase between two two-dimensional 
defects stacked parallel in a four-dimensional ambient spacetime.
Different UV parameters are turned on for different defects.
We study the system 
in the quenched strong coupling limit, using holography with probe brane approximation, 
realized explicitly through the D3/D5 system. 
This coupled phase is holographic dual to the presence of spontaneous symmetry breaking 
of the individual ultraviolet flavor symmetries associated with the double heterostructure 
of the defect layers. 
We characterize this solution by its infrared geometric data, 
and present the numerical result showing a first-order phase transition 
between the asymmetrically coupled phase and the more mundane decoupled phase.

}

\begin{document}
\maketitle

\section{Introduction} 

The $AdS$/$CFT$ correspondence\cite{Maldacena:1997re,Witten:1998qj,Gubser:1998bc,Aharony:1999ti} 
has provided the physics community with an unique perspective 
in the attempt to understand the various phases of strongly coupled quantum field theory, 
and together with 
the probe-brane extension thereof\cite{Karch:2002sh,Karch:2000gx,DeWolfe:2001pq}, 
it has been providing the theoretical playground that 
renders possible  toy-model building for a wide 
domain of interests, 
ranging from the quark-gluon plasma\cite{CasalderreySolana:2011us,Schafer:2009dj}, 
unitary Fermi gas\cite{Son:2008ye,Balasubramanian:2008dm}, 
to high-$T_c$ superconductor\cite{Hartnoll:2008kx,Hartnoll:2009ns}. 

Of special relevance to this work are the various deformations associated with a probe-brane 
embedding submanifold. Such considerations are crucial in the construction of the  Saki-Sugimoto 
D4/D8 model\cite{Sakai:2004cn}, dual to the holographic QCD, and also fruitful in identifying various phases of the strongly coupled quantum field theory dual to the D3/D7 model\cite{Karch:2002sh}. 
Nevertheless, in the aforementioned  D4/D8 and D3/D7 systems, the probe-brane embedding 
submanifold actually spans the whole four-dimensional external spacetime of the field theory, 
rendering possible only the deformation associated with the internal space. 
This of course is due to the feature of the physical system under study, where the quark sector 
roams freely together with the gluon sector. 
However, the idea of deformation of submanifolds, and possible interesting physical systems, 
are both admissible to more general consideration: Focused on the D3/D5 system, 
where the probe-brane system describes the defect sector confined into a two-dimensional plane, 
a nontrivial deformation of the submanifold for both the external and internal embedding is possible.
In fact, by stacking two such defects parallel, 
we show that, 
even when the two defect contents are in general different in their UV parameters,
hence holographic dual to double heterostructure, 
there still exists an asymmetrically connected submanifold solution for the probe brane system.
This configuration is
dual to a coupled phase in the field theory side. 
We also find that such an asymmetric configuration
is thermodynamically favorable upon some mild detuning of UV data, 
and there is a first-order transition from  this asymmetrically 
coupled phases to the more mundane decoupled phase, with the increase 
of the difference of their UV parameters.

The organization of the paper is as follows:
In Section 2 we first review the notion of the D7 and D5 embeddings in $AdS_5\times S^5$,
which also provides the mathematical setting. We also introduce the notion of single-sheeted, 
also later called decoupled phase, for the rest of the work.  
After that we present the calculation subtleties associated with 
the construction of the asymmetric joint-sheeted, 
also called nontrivial coupled phase.
In Section 3 we present the numerical results, 
showing a first-order transition between these two aforementioned phases. 


\section{Methods \label{sec_result}} 

In this section, we introduce the notion used by Graham and Karch\cite{toappear} 
for setting up our convention. They consider 
submanifolds asymptotic to $AdS_{k+1}\times S^{l}$ inside $AdS_{n+1}\times S^{m}$.
Specialized to the D3/D5 embedding, $k=3, l=2, n=4, m=5$, this probe brane system
is  holographic dual to 
two localized defects\cite{Karch:2002sh,DeWolfe:2001pq}, 
with quark content of massive $\mathcal{N}=2$ $N_f$-favor hypermultiplets in the fundamental representation, 
inserted into 
the background of  $\mathcal{N}=4$ $U(N_c)$ super-Yang-Mills gauge field theory, 
in the $N_c \gg 1$, large 't Hooft coupling limit. 
The two defects are separated by $\Delta x$ in the common transverse direction, 
and each sector is characterized by its own quark mass $m_L$ or $m_R$ in the UV Lagrangian. 
We will construct the phase diagram by tuning the dimensionless ratios, $m_L \Delta x$ and $m_R \Delta x$.

\subsection{Submanifold Extension: mathematical settings}
The mathematical setting of this article is as the following: 
Considering the background spacetime being a product manifold $X \times K$, 
where $X$ is a $n+1$ dimensional asymptotically hyperbolic manifold 
with $n$ dimensional boundary thereof, $M \equiv \partial X $,
and $K$ is a $m$ dimensional compact manifold.  
The metric of $X \times K$ is parametrized using the standard 
Fefferman-Graham form\cite{Fefferman:1985}:
\begin{equation}
g=g_{+} + g_{K} = \frac{dr^2+\bar{g}_{r}}{r^2}+g_{K},
\end{equation}
with boundary $M$ located at $r_{min}=0$, hereafter referred as the UV end. 
The minimal submanifold is denoted as $Z \subset X \times K$, 
with boundary $\partial Z = N \times S \subset M \times K$. 
We can parametrize $M$ as $(x^{\alpha},u^{\alpha'})$,  $K$ as $(s^{A},t^{A'})$, 
such that: 1) The boundary of the submanifold 
$N \times S$ is given by $u^{\alpha'}\rightarrow 0$, 
$t^{A'}\rightarrow 0$,
and 2) $t$ and $u$ variables are ``orthogonal'' to the boundary, as in the following sense:
\begin{equation}
\bar{g}_{\alpha,\alpha'}|_{r=0,u=0}=g_{A A'}|_{t=0}=0.
\end{equation}
In this article,
we focus on submanifolds asymptotic to $AdS_{k+1}\times S^l$, embedding into $AdS_{n+1}\times S^{m}$.
We will choose to embed $S^{l}$ into $S^{m}$ as follows:
\begin{equation}
\label{eq:spheremetric}
ds^2_{S^m}=
d\theta^2
+\cos^2\theta d\Omega^2_l
+\sin^2\theta d\Omega^2_{m-l-1},
\end{equation}
with $S^{l}$ sitting at the equator parametrized by the $\theta$ embedding function. This ansatz 
corresponds to the simplest symmetry-breaking pattern associated with the dual defect field theory. 

Given the above parametrization, 
let us first consider the single-sheeted (later also called decoupled) phase of the minimal submanifold. 
In this phase, the minimal submanifold is not only asymptotic to $AdS_{k+1}\times S^{l}$,
but when it extends into the bulk,
the only change is the internal sphere radius, controlled by the $\theta$ embedding function, 
being endowed with nontrivial radial dependence, $\theta=\theta(r)$.
All external-space embeddings are constant, 
hence no ``bending'' of the probe external location occurs along the radial direction. 
In fact, notice that radial dependence is actually the most generally allowed dependence 
if we insist on preserving 
the defect-translational invariance, as well as
$S^{l}$ and $S^{m-l-1}$ isometries for the dual field theory.
With the standard Poincar\'e patch for $AdS_{n+1}$ and static gauge for $AdS_{l+1}$,
the area of the submanifold is then given by:
\begin{equation}
S = \int dr 
\frac{\cos^{l}\theta}{r^{k+1}}
\sqrt{1+ \alpha^2 r^2(\theta')^2},
\end{equation}
where 
$\alpha$ is a generalization introduced 
to account for the possible difference between the 
curvature radii of internal compact space $S^{m}$ and 
the external $AdS_{k+1}$ space, $\alpha\equiv(R_{S^m}/R_{AdS_{k+1}})$. 
 $\alpha = 1$ in 
$AdS_{5}\times S^{5}$, 
the standard near-throat limit of supergravity background generated by from the D3 brane. 
Given that the  D3/D5 system is our primary concern, hereafter we commit ourselves to the $\alpha = 1$ case.
Also we have set $R_{AdS_{k+1}}$ to $1$.
The usual variational method yields the equation of motion for $\theta(r)$ as:
\begin{equation}
\label{eom:singlesheeted}
\theta'' =- l \left(\frac{1}{r^2} + (\theta')^2  \right) \tan\theta
+ ( \frac{-1+k}{r} ) \theta' + k r (\theta')^3.
\end{equation}
Let's first consider the special case of  the D3/D7 system, 
where the D7 probe brane is asymptotic to $AdS_5\times S^3$ ($k=4, l=3$):
This submanifold is completely filling the external space of the background spacetime,
and we can easily check that we then have an one-parameter family of embeddings,
$\theta=\arcsin(mr)$,
with $m$ being the free parameter. $\theta$ is the so-called slipping mode in the literature. 
The same solution also holds for the D3/D5 system, with the probe brane embedding submanifold
asymptotic to  $AdS_4\times S^2$ ($k=3, l=2$).
Such a simple analytical result is connected with the supersymmetry of 
the probe brane system\cite{Brunner:1998jr}.
In more general systems, however,
one can only observe that $\theta$ being zero is still a solution, 
which extends into the one-parameter family of solutions by 
turning on the slipping mode with the initial slope, 
$\theta(r)=mr+\mathcal{O}(r^2)$. 
But one will need to numerically integrate out the equation to complete the profile.

However, we are naturally more interested in configurations with the external embedding being nontrivial, 
specifically the joint-sheeted configuration, corresponding to two defect stacked parallel to each other, 
with the submanifolds bending towards each other and eventually smoothly joined in the radial direction.
An example of this connected configuration is
the Wilson line used by Maldacena\cite{Maldacena:1998im} to compute the quark-antiquark force: 
Near the boundary the submanifold is asymptotic to $AdS_2$ times a point. 
However, deep inside the bulk, the submanifold is connected into a U-shaped configuration.
Therefore, 
in this article, we parametrize such a connected submanifold as being the union of two different branches,
$\partial Z = (N_1\times S_1) \bigcup (N_2 \times S_2) $, 
with each branch chosen to be $N_i \times S_i = R^k \times S^l$.
As mentioned before, to preserve the dual field theory symmetries associated with the symmetries of 
both the defect space and the transverse space, 
only radial dependence is admissible for the embedding functions. 
For simplicity, 
instead of the simple internal global symmetry breaking pattern chosen by Eq.(\ref{eq:spheremetric}),
we will also restrict ourselves by only turning on one nontrivial external embedding function, 
hereafter named $x(r)$.
Still working in the standard Poincar\'e patch for $AdS_{n+1}$ and static gauge for $AdS_{l+1}$,
the area of the submanifold is  modified as:
\begin{equation}
S=\int dr \frac{cos^l\theta}{r^{k+1}}\sqrt{1+(x')^2+   r^2(\theta')^2},
\end{equation}
with the equations of motion given by:
\begin{align}
x'' =& \frac{(1+k)}{r} (x')^3+ x' \left(\frac{1+k}{r}+ k r   (\theta')^2\right); \nonumber \\
\theta'' =& 
-l \left( \frac{1+(x')^2}{ r^2} + (\theta')^2 \right) \tan\theta
+ \frac{-1+k+(1+k)(x')^2}{r} \theta' + k r (\theta')^3.
\label{eom:joinsheeted}
\end{align}

However, unlike the previous single-sheeted phase, 
these equations admit no analytical solution at least to us, 
and hence we are forced to resort to numerical methods for computing the profile.


\begin{figure}[h]
     \centering
     \begin{subfigure}[b]{0.3\textwidth}
     \includegraphics[width=\textwidth]{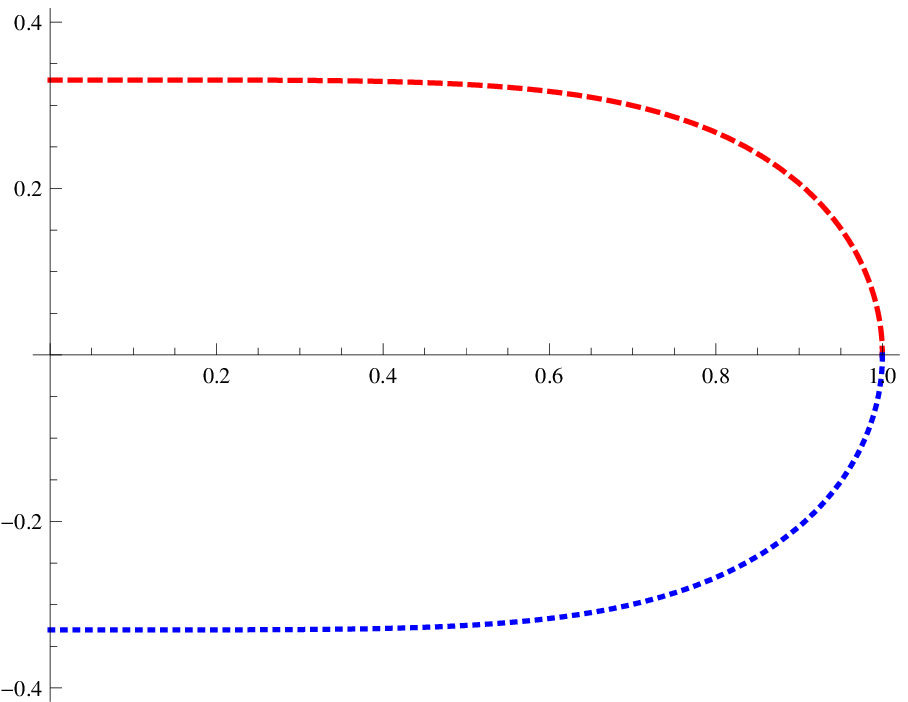}
     \caption{$x(r)_{symmetric}$}
     \label{fig:symx}
      \end{subfigure}
     \begin{subfigure}[b]{0.3\textwidth}
     \includegraphics[width=\textwidth]{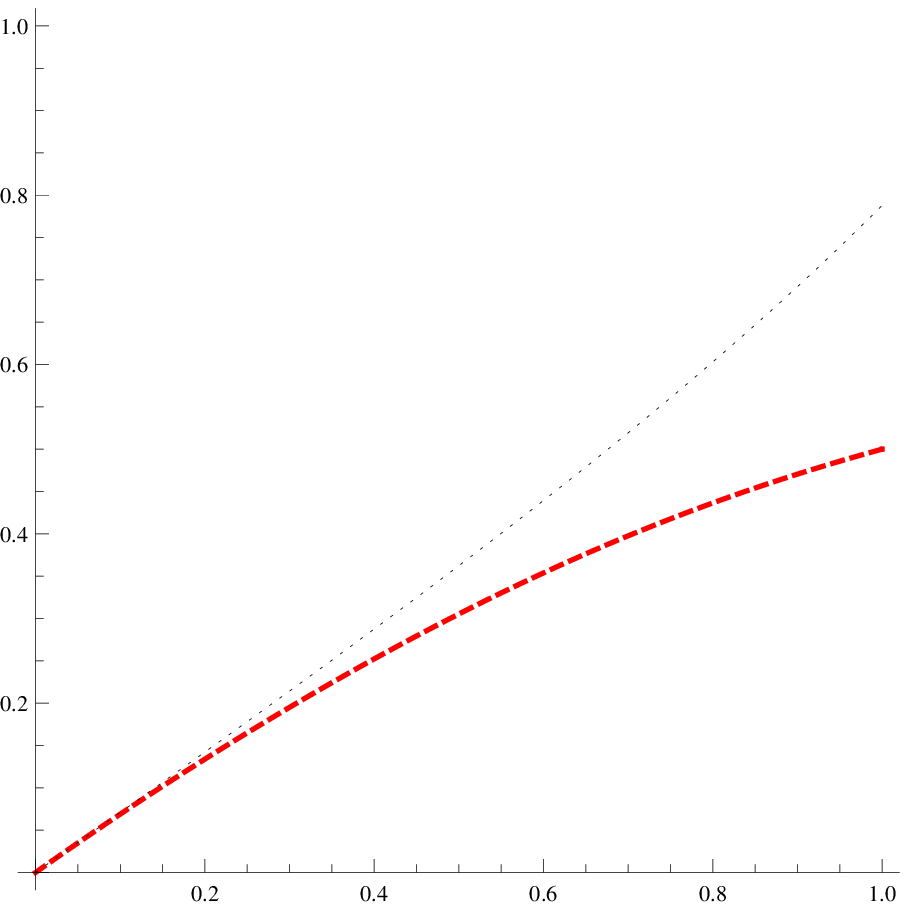}
      \caption{$\theta(r)_{symmetric}$}
     \label{fig:symtheta}
      \end{subfigure}
\caption{
A typical result for a symmetrically connected configuration. Within
the $r$-parametrization, this coupled configuration 
is expressed by two separate but identical 
branches joining smoothly at the turning point $r_{max}$, normalized to $1$ in above. 
In Fig.\ref{fig:symtheta}, the dotted line stands for 
the asymptotically decoupled configuration with the same UV parameter,
the asymptotic mass term 
$m_{asymptotics}\equiv \theta'(r)\big|_{r\rightarrow 0}.$ 
 Notice we use the translational invariance to set $x(r_{max})=0$.
}
\label{fig:symplot}
\end{figure}

\begin{figure}[h]
     \centering
     \begin{subfigure}[b]{0.3\textwidth}
     \includegraphics[width=\textwidth]{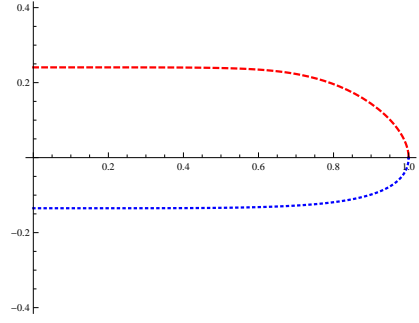}
     \caption{$x(r)_{asymmetric}$}
     \label{fig:asymx}
      \end{subfigure}

     \begin{subfigure}[b]{0.3\textwidth}
     \includegraphics[width=\textwidth]{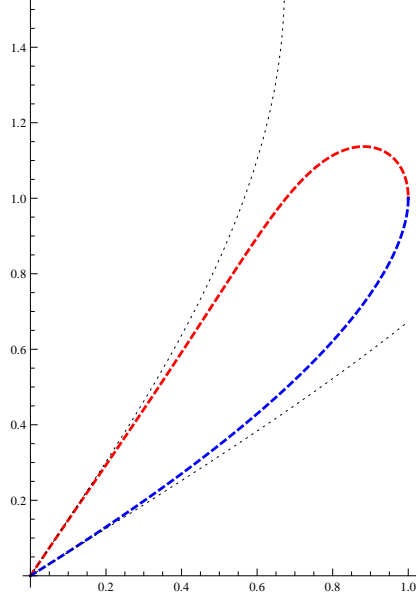}
      \caption{$\theta(r)_{asymmetric}$}
     \label{fig:asymtheta}
      \end{subfigure}
\caption{
A typical result for an asymmetrically connected configuration. Within
the $r$-parametrization, this coupled configuration 
is expressed by two separate and different 
branches joining smoothly at turning point $r_{max}$, normalized to $1$ in above. 
In Fig.\ref{fig:asymtheta}, the dotted lines stand for 
the corresponding asymptotically decoupled configuration for each branch,
with different mass terms on the different side of the lobe. We
can generate such graph within the $x$-parametrization from the IR end, and
cascade with the $r$-parametrization when approaching the UV end.
Notice we use the translational invarience to set $x(r_{max})=0$.
}\label{fig:asymplot}
\end{figure}

\subsection{The Regularity Constraint and the Cascaded Integration Scheme}

To solve the embedding equations Eq.(\ref{eom:joinsheeted}), 
the boundary conditions, required to generate the joint-sheeted configurations, 
are given by  requiring the connection of the two branches inside the bulk space being smooth. 
This however 
renders $x'(r)\big|_{r\rightarrow r_{max}}$ divergent and further 
denies us control over the boundary specification at $r_{max}$, 
the turning point of the joint-sheet submanifold. 
Nevertheless, this seeming difficulty can be resolved by a closer inspection of the 
regularity requirement: 
Notice that the Lagrangian density is cyclic in the $x$-parameter, hence conserving the conjugated momentum thereof,
\begin{equation}
\Pi_x\equiv \frac{\partial L}{\partial x'} = 
\frac{cos^l\theta}{r^{k+1}}\frac{x'}{\sqrt{1+(x')^2+ r^2(\theta')^2}}.
\end{equation}
Therefore, we can express $x'$ in term of this integral of motion $\Pi_x$,
\begin{equation}
x'=\pm \sqrt{\frac{1+ r^2 (\theta')^2}{ \left(\ \frac{cos^l\theta}{\Pi_x r^{k+1}} \right)^2 -1 }}.
\end{equation}
The regularity requirement therefore translates into the following two scenarios:
First, we have the denominator equal to zero, providing a condition relating all the parameters at the IR end:
\begin{equation}
\left[ \left(\ \frac{cos^l\theta}{\Pi_x r^{k+1}} \right)^2 -1 \right] 
\xrightarrow{r\rightarrow r_{max}}  0 ;
\end{equation}
Second, we have the numerator equal to infinity, signaling at the IR end sitting not only a divergent $x'(r_{max})$ 
but also a divergent $\theta'(r_{max})$ as well:
\begin{equation}
\frac{d\theta}{dr} \xrightarrow{r\rightarrow r_{max}} \infty  .
\end{equation}
One may expect the first scenario to be more relevant, 
given the smoother behavior of $\theta(r)\big|_{r\rightarrow r_{max}}$.
But further analysis, by expanding the submanifold in power series given the regular behavior of $\theta(r)$ around $r_{max}$,  reveals that the first scenario is actually too restrictive.
It constrains the two branches of the submanifold to be exactly the same, 
ending with only symmetric joint-sheet configurations.
Therefore, 
to include the most general, asymmetric joint-sheeted configuration, the second scenario and 
hence singular behavior of $\theta(r)$ around $r_{max}$ will need to be considered. 

To numerically generate the solution given by the second scenario for the boundary conditions, 
we adopt the following cascaded integration scheme:
First, around $r\sim r_{max}$, 
we choose to parametrize the solution in terms of the $x$-variable, 
with the equations of motion given by:
\begin{align}
r''&= - \frac{1+k}{r} (1+(r')^2) - k r (\theta')^2;
\\ 
\theta''&= - l \left( \frac{1+(r')^2}{r^2} + (\theta')^2 \right) \tan\theta  - 2 \frac{r'}{r} \theta'.
\end{align}
Choosing the turning point to be at $x_0=0$, the solution is uniquely determined by the ``IR'' data: 
$\{r(x_0) \equiv 1, r'(x_0)\equiv 0,\theta(x_0), \theta'(x_0)\}$, located at $r_{max}$. 
Therefore, the entire solution family are indeed generated by 4 parameters: 
$\{\theta(x_0), \theta'(x_0),r(x_0),x_0\}$, with the later 2 generated from dilatation and translation symmetry.
Given the equations are regular in terms of the $x$-variable,
we perform the numerical integration to a predetermined intermediate point away from the boundary. 
However, given the equations  is singular in terms of $x$-variable at the boundary,
to approach the boundary with more numerical stability,  
we will then switch over to the $r$-parametrization before carrying out the integration to the boundary.
With such a cascaded scheme, we can numerically 
find the most general, asymmetric connected configuration.

\subsection{Extraction Scheme}

The great virtue of the probe brane approximation is that various physical quantities of interest
can be obtained without re-solving the Einstein equation with the probe source, 
as long as those quantities can be calculated using the free energy and thermodynamical relations, 
as already explained in \cite{Karch:2008uy}. 
Therefore, to determine the phase diagram, 
we can follow this tenet and simply compute the free energy 
using the on-shell action for the given configuration,
and determine which phase, 
single-sheeted (also called decoupled) or 
joint-sheeted (also called coupled), is thermodynamically favorable. 
However, the boundary divergence of $AdS$ space renders the first step rather laborious: 
the canonical approach is to deploy the holographic renormalization\cite{deHaro:2000xn}, 
which carefully reconstructs the diffeomorphism-invariant counterterms 
by examining the divergent structure associated with the 
tentative cut-off plane, dual of the UV regulator for field theories. 
However, in this study, given that we eventually only focus on the difference of free energies between 
the decoupled and coupled configurations, 
we will adopt the following extraction scheme (also known as background subtraction): given any coupled configuration,
we first construct the dual disconnected configuration with the same UV parameter $m_{asymptotic}$. 
Then the difference of these two configurations is, by construction, vanishing near the boundary, 
since all coupled configurations are asymptotic to the decoupled solution in the UV region 
(See Fig.\ref{fig:symplot}-\ref{fig:asymplot}).

The rationale behind our extraction scheme is that 
the to-be-constructed holographic counterterms can only depend on the UV behaviors but not IR physics,
which is already present in the decoupled configuration once the only relevant information, 
$m_{L}$ or $m_{R}$ is extracted. 
This scheme is also more in tune with the pre-Wilsonian 
renormalization philosophy, ``sweeping under the rug'', 
that no divergence should be present if every physical prediction 
is expressed with physical quantities.

In practice, given we can only work with the numerically generated configuration, 
there is a potential caveat associated with such an extraction scheme: 
the UV parameter $m_{L/R}$  is located at the singular point of the equations of motion, 
and we only extract this information numerically up to a small cut-off distance due to the inevitable numerical instability. 
Such a seemingly simple numerical recipe can be subject to more elaborate modification:
One can construct the analytic solution expanded around the singular point, 
and extract the parameters by fitting at a point away from boundary, with now more numerical control.
However, given the phase transition we find is of first order in nature, 
and in the precision we are working on,
the difference thus introduced is found to be numerically negligible with no qualitative change of the conclusion. 


\begin{figure}[h]
\centering
\includegraphics[width=0.5\textwidth]{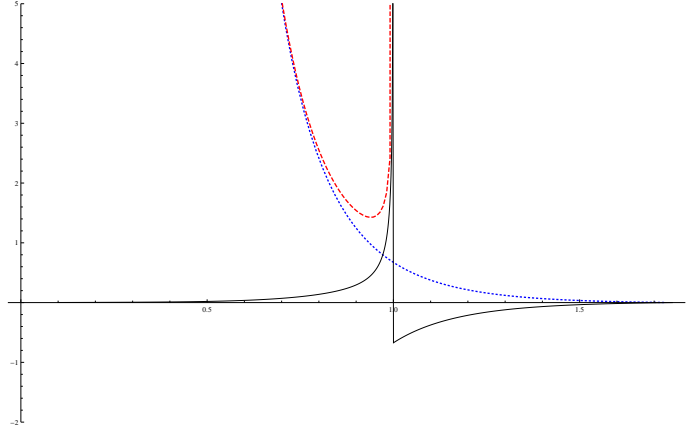}
\caption{
A typical extraction result of comparing the Lagrangian (submanifold volume
density) with $r$, the AdS radial direction: 
Red(dashed) line, a typical coupled configuration; 
Blue(dotted) line, the dual decoupled configuration; 
Black(thin) line, the difference between the previous two, with the signed area being the free energy difference. 
One can observe the divergence term is indeed subtracted out when approaching
the boundary, 
and the difference is only due to the back reaction inside the
bulk. Notice in this specific case the decoupled configuration 
can extend deeper into the bulk even after the coupled counterpart 
already terminates ($r_{max} = 1$ by normalization). 
The reverse situation also is possible, where the decoupled surface
can terminate before the connected one.
}
\label{fig:lag}
\end{figure}


\section{Phase Diagram}

\subsection{Symmetric Coupled Phase}
\begin{figure}[h]
\centering
\includegraphics[width=0.5\textwidth]{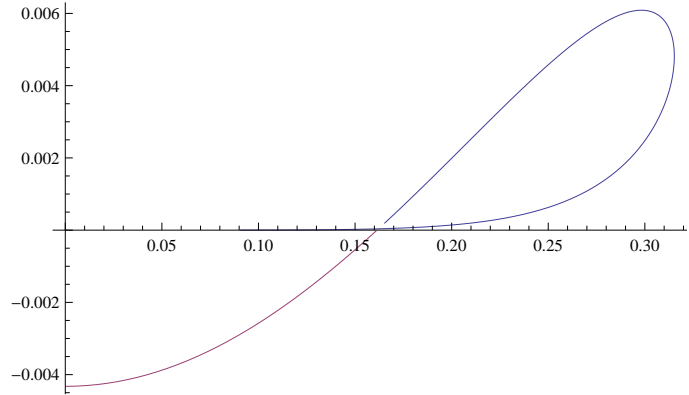}
\caption{
Indication of the first order phase transition of symmetrically coupled 
submanifold
as $\xi_{normalized}$ changes. 
The vertical axis shows the difference of surface area
between the coupled configuration and its decoupled counterpart; 
the horizontal axis is the characteristic label of the connected surface defined in Eq.(\ref{mphysical})
}
\label{fig:symphasetransition}
\end{figure}

We first restrict to the symmetric connected configurations only.
Fig.\ref{fig:symphasetransition} shows
the on-shell area difference, which is also the free energy difference, 
between the symmetric coupled and decoupled
configurations as a function of the normalized mass term $\xi_{normalized}$:
\begin{align}
\label{mphysical}
\xi_{normalized} = m  \Delta x, 
\end{align}
with
\begin{align}
&m = \theta'(r_{min})  \nonumber \\
&\Delta x = x(r_{min})\big|_{upper-branch} -  x(r_{min})\big|_{lower-branch}.
\end{align}
For every value of $\xi_{normalized}$ we find two coupled solutions.
At very low
$\xi_{normalized}$, 
one of the symmetric coupled configuration is the thermodynamically favored phase, compared with its decoupled counterpart(with negative free energy difference);
as $\xi_{normalized}$ increases, 
the area difference begins to shrink, and finally a
first order phase transition is reached 
when $\xi_{normalized}$ reach the value of $0.165$,
after which the decoupled dual becomes more favorable
ones; however, as $\xi_{normalized}$ 
keeps increasing and eventually above $0.315$, the coupled configuration disappears entirely, 
and only the decoupled phase exists as the only allowed solution for the minimal submanifold.

\subsection{Asymmetric Coupled Phase}
\begin{figure}[h]
     \centering
     \begin{subfigure}[b]{0.5\textwidth}
     \includegraphics[width=\textwidth]{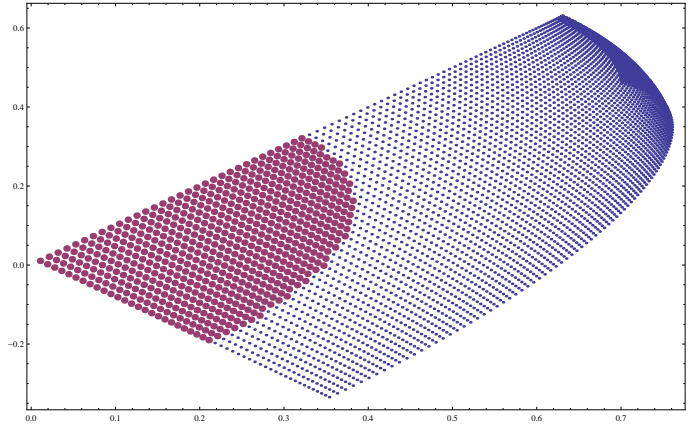}
     \caption{Phase Diagram for Asymmetrically Coupled Configuration}
     \label{fig:asymcaseflat}
      \end{subfigure}

     \begin{subfigure}[b]{0.5\textwidth}
     \includegraphics[width=\textwidth]{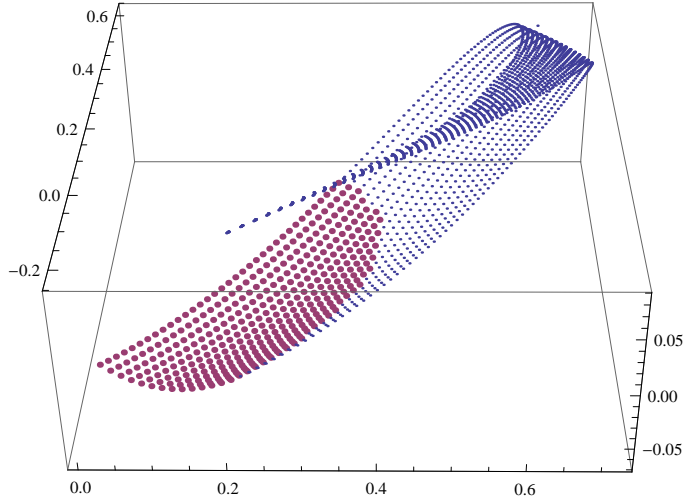}
      \caption{
A snapshot of numerical scanning of the 
free energy difference with constant $\theta'(r_{max})$-curves
}
     \label{fig:asymcase3d}
      \end{subfigure}
\caption{
Indication of the first order phase transition of asymmetrically coupled 
submanifold
as $\xi_{L}$ and $\xi_{R}$ are varied. 
Given the solution is enumerated with the IR data, 
$\theta(r_{max})$ and $\theta'(r_{max})$, 
we perform the numerical scan by 
constructing the constant $\theta'(r_{max})$-curve and 
slowly filling up the entire phase space, where in the red (heavy-shaded) region
the coupled phase dominates.
Notice the upper corner is present due to our scanning procedure: 
The solution is scanned by varying the IR data, 
but the phase diagram is labeled by the dimensionless parameters back-constructed from UV properties of the 
obtained solution. This leads to the same upper lope as in Fig.\ref{fig:symphasetransition}.
}\label{fig:asymphasetransition}
\end{figure}

Using the cascade evolution scheme, we can also compute the phase diagram for 
the asymmetric connected configuration. Fig. \ref{fig:asymphasetransition} 
shows the phase diagram of the first order phase transition for the minimal submanifold between
the asymmetric coupled and decoupled configurations as a
function of two physical parameters:
\begin{align}
\xi_{i} = m_{i} (\Delta x ) , i \in \{L,R\} ;
\end{align}
with
\begin{align}
&m_{i} = \theta'_i(r_{min})  \nonumber \\
&\Delta x_i = \big| x_i(r_{min})\big|, i \in \{L,R\},
\end{align}
where we use $L$ and $R$ to denote the different branch under study.
Notice that due to the reflection symmetry of the parameter space along
the axes of both $\xi_L = \xi_R$ and $\xi_L = - \xi_R$, 
we only need to scan the portion of Fig. \ref{fig:asymphasetransition}.
In Fig.\ref{fig:asymcase3d}, the partial scanning of the free energy density difference 
is also shown,  which explicitly demonstrates that the phase transition is also of the first-order.




\section{Conclusion and Future Research} 

In this article we investigate the submanifolds asymptotic to $AdS_4 \times S^2$ 
embedded into $AdS_5\times S^5$. We restrict to the largest unbroken residual symmetry.
We find the asymmetric joint-sheeted configuration, 
with both external embedding and internal embedding radially deformed.
These are dual to the coupled phase between two two-dimensional 
defects stacked parallel, with different UV parameters.
While similar phases have been studied previously in the probe brane systems, 
what is novel about our work is that
the asymmetric probe brane system is holographic dual to  
the double heterostructure in condensed matter systems,
with different UV data in the two layers.
The asymmetrically coupled phase is 
holographic dual to the spontaneous symmetry breaking of the original decoupled 
ultraviolet
$U(N_{f})_L\times U(N_{f})_R$ flavor symmetries, 
associated with flavor rotation for individual defect content, 
into the diagonal subgroup $U(N_{f})$ due to the non-vanishing condensate developed at infrared, 
$\langle \bar{\psi}^{i}_{a} {(e^{\int A})^a}_{b} \psi^{j,b} \rangle$.
This order parameter has also been studied in \cite{Aharony:2008an} for the chiral condensate of 
Saki-Sugimoto model.
Solutions are shown completely parametrized by the infrared geometric data,
being the IR value and slope of the internal deformation,
upon which the entire profile is numerically constructed by the 
cascading integration scheme we adopted in this paper. 
Aimed with the complete numerical solution, 
we map out the phase diagram between this asymmetrically coupled phase and
the competing mundane decoupled phase, and we find that the coupled phase is dominating 
at zero temperature, even when the UV parameters are mildly detuned, and with large enough difference 
the system undergoes a first-order phase transition to the decoupled phase. 

Notice that more general deformations of the submanifolds are possible, 
corresponding to finer symmetry breaking pattern in the dual field theory side.
Such details present no challenge for profile construction using the cascading integration scheme.
However the corresponding phase diagram is computationally difficult to enumerate:
For example, in the cases of deforming two internal embedding functions, 
{\it ie.} $\theta=\theta(r)$ and $\psi=\psi(r)$, 
we find that this solution can be constructed with 4 degrees of freedom, 
being their IR values and slopes of the two internal coordinates.
The phase diagram will be 4 dimensional, which present itself an expensive numerical barrier. 
Tracing out this space to locate the transition boundary and 
hence its transition nature will be left for future investigation.


\section*{Acknowledgments}
HC will like to thank Andreas Karch and Robin Graham for sharing their work and 
various essential insights during various discussion of the project. 
This work has been supported
in part by the U.S. Department of Energy under Grant No. DE-FG02-96ER40956.

\bibliography{ref}

\providecommand{\href}[2]{#2}\begingroup\raggedright\begin{thebibliography}{10}

\bibitem{Maldacena:1997re}
J.~M. Maldacena, {\it {The Large N limit of superconformal field theories and
  supergravity}},  {\em Adv.Theor.Math.Phys.} {\bf 2} (1998) 231--252,
  [\href{http://xxx.lanl.gov/abs/hep-th/9711200}{{\tt hep-th/9711200}}].

\bibitem{Witten:1998qj}
E.~Witten, {\it {Anti-de Sitter space and holography}},  {\em
  Adv.Theor.Math.Phys.} {\bf 2} (1998) 253--291,
  [\href{http://xxx.lanl.gov/abs/hep-th/9802150}{{\tt hep-th/9802150}}].

\bibitem{Gubser:1998bc}
S.~Gubser, I.~R. Klebanov, and A.~M. Polyakov, {\it {Gauge theory correlators
  from noncritical string theory}},  {\em Phys.Lett.} {\bf B428} (1998)
  105--114, [\href{http://xxx.lanl.gov/abs/hep-th/9802109}{{\tt
  hep-th/9802109}}].

\bibitem{Aharony:1999ti}
O.~Aharony, S.~S. Gubser, J.~M. Maldacena, H.~Ooguri, and Y.~Oz, {\it {Large N
  field theories, string theory and gravity}},  {\em Phys.Rept.} {\bf 323}
  (2000) 183--386, [\href{http://xxx.lanl.gov/abs/hep-th/9905111}{{\tt
  hep-th/9905111}}].

\bibitem{Karch:2002sh}
A.~Karch and E.~Katz, {\it {Adding flavor to AdS / CFT}},  {\em JHEP} {\bf
  0206} (2002) 043, [\href{http://xxx.lanl.gov/abs/hep-th/0205236}{{\tt
  hep-th/0205236}}].

\bibitem{Karch:2000gx}
A.~Karch and L.~Randall, {\it {Open and closed string interpretation of SUSY
  CFT's on branes with boundaries}},  {\em JHEP} {\bf 0106} (2001) 063,
  [\href{http://xxx.lanl.gov/abs/hep-th/0105132}{{\tt hep-th/0105132}}].

\bibitem{DeWolfe:2001pq}
O.~DeWolfe, D.~Z. Freedman, and H.~Ooguri, {\it {Holography and defect
  conformal field theories}},  {\em Phys.Rev.} {\bf D66} (2002) 025009,
  [\href{http://xxx.lanl.gov/abs/hep-th/0111135}{{\tt hep-th/0111135}}].

\bibitem{CasalderreySolana:2011us}
J.~Casalderrey-Solana, H.~Liu, D.~Mateos, K.~Rajagopal, and U.~A. Wiedemann,
  {\it {Gauge/String Duality, Hot QCD and Heavy Ion Collisions}},
  \href{http://xxx.lanl.gov/abs/1101.0618}{{\tt arXiv:1101.0618}}.

\bibitem{Schafer:2009dj}
T.~Schafer and D.~Teaney, {\it {Nearly Perfect Fluidity: From Cold Atomic Gases
  to Hot Quark Gluon Plasmas}},  {\em Rept.Prog.Phys.} {\bf 72} (2009) 126001,
  [\href{http://xxx.lanl.gov/abs/0904.3107}{{\tt arXiv:0904.3107}}].

\bibitem{Son:2008ye}
D.~Son, {\it {Toward an AdS/cold atoms correspondence: A Geometric realization
  of the Schrodinger symmetry}},  {\em Phys.Rev.} {\bf D78} (2008) 046003,
  [\href{http://xxx.lanl.gov/abs/0804.3972}{{\tt arXiv:0804.3972}}].

\bibitem{Balasubramanian:2008dm}
K.~Balasubramanian and J.~McGreevy, {\it {Gravity duals for non-relativistic
  CFTs}},  {\em Phys.Rev.Lett.} {\bf 101} (2008) 061601,
  [\href{http://xxx.lanl.gov/abs/0804.4053}{{\tt arXiv:0804.4053}}].

\bibitem{Hartnoll:2008kx}
S.~A. Hartnoll, C.~P. Herzog, and G.~T. Horowitz, {\it {Holographic
  Superconductors}},  {\em JHEP} {\bf 0812} (2008) 015,
  [\href{http://xxx.lanl.gov/abs/0810.1563}{{\tt arXiv:0810.1563}}].

\bibitem{Hartnoll:2009ns}
S.~A. Hartnoll, J.~Polchinski, E.~Silverstein, and D.~Tong, {\it {Towards
  strange metallic holography}},  {\em JHEP} {\bf 1004} (2010) 120,
  [\href{http://xxx.lanl.gov/abs/0912.1061}{{\tt arXiv:0912.1061}}].

\bibitem{Sakai:2004cn}
T.~Sakai and S.~Sugimoto, {\it {Low energy hadron physics in holographic QCD}},
   {\em Prog.Theor.Phys.} {\bf 113} (2005) 843--882,
  [\href{http://xxx.lanl.gov/abs/hep-th/0412141}{{\tt hep-th/0412141}}].

\bibitem{toappear}
C.~Graham and A.~Karch {\em in preparation}.

\bibitem{Fefferman:1985}
C.~R.~G. C.~Fefferman, {\it Conformal invariants},  {\em Ast\'erisque, Numero
  Hors Serie} {\bf 95} (1985).

\bibitem{Brunner:1998jr}
I.~Brunner, A.~Hanany, A.~Karch, and D.~Lust, {\it {Brane dynamics and chiral
  nonchiral transitions}},  {\em Nucl.Phys.} {\bf B528} (1998) 197--217,
  [\href{http://xxx.lanl.gov/abs/hep-th/9801017}{{\tt hep-th/9801017}}].

\bibitem{Maldacena:1998im}
J.~M. Maldacena, {\it {Wilson loops in large N field theories}},  {\em
  Phys.Rev.Lett.} {\bf 80} (1998) 4859--4862,
  [\href{http://xxx.lanl.gov/abs/hep-th/9803002}{{\tt hep-th/9803002}}].

\bibitem{Karch:2008uy}
A.~Karch, A.~O'Bannon, and E.~Thompson, {\it {The Stress-Energy Tensor of
  Flavor Fields from AdS/CFT}},  {\em JHEP} {\bf 0904} (2009) 021,
  [\href{http://xxx.lanl.gov/abs/0812.3629}{{\tt arXiv:0812.3629}}].

\bibitem{deHaro:2000xn}
S.~de~Haro, S.~N. Solodukhin, and K.~Skenderis, {\it {Holographic
  reconstruction of space-time and renormalization in the AdS / CFT
  correspondence}},  {\em Commun.Math.Phys.} {\bf 217} (2001) 595--622,
  [\href{http://xxx.lanl.gov/abs/hep-th/0002230}{{\tt hep-th/0002230}}].

\bibitem{Aharony:2008an}
O.~Aharony and D.~Kutasov, {\it {Holographic Duals of Long Open Strings}},
  {\em Phys.Rev.} {\bf D78} (2008) 026005,
  [\href{http://xxx.lanl.gov/abs/0803.3547}{{\tt arXiv:0803.3547}}].

\end{thebibliography}\endgroup
\bibliographystyle{JHEP}

\end{document}